# From LiDAR to Underground Maps via 5G - Business Models Enabling a System-of-Systems Approach to Mapping the Kankberg Mine

Markus Borg, Thomas Olsson
{markus.borg, thomas.olsson}@sics.se
Software and Systems Engineering Laboratory
RISE SICS AB
Lund, Sweden

John Svensson
john.svensson@boliden.se
Linköping University
Boliden AB
Boliden, Sweden

## Abstract
With ever-increasing productivity targets in mining operations, there is a growing interest in mining automation. The PIMM project addresses the fundamental challenge of network communication by constructing a pilot 5G network in the underground mine Kankberg. In this report, we discuss how such a 5G network could constitute the essential infrastructure to organize existing systems in Kankberg into a system-of-systems (SoS). In this report, we analyze a scenario in which LiDAR equipped vehicles operating in the mine are connected to existing mine mapping and positioning solutions. The approach is motivated by the approaching era of remote controlled, or even autonomous, vehicles in mining operations. The proposed SoS could ensure continuously updated maps of Kankberg, rendered in unprecedented detail, supporting both productivity and safety in the underground mine. We present four different SoS solutions from an organizational point of view, discussing how development and operations of the constituent systems could be distributed among Boliden and external stakeholders, e.g., the vehicle suppliers, the hauling company, and the developers of the mapping software. The four scenarios are compared from both technical and business perspectives, and based on trade-off discussions and SWOT analyses. We conclude our report by recommending continued research along two future paths, namely a closer cooperation with the vehicle suppliers, and further feasibility studies regarding establishing a Kankberg software ecosystem.

## 1 Introduction
The global demand for raw materials is increasing, thus mining operations are producing close to their capacity limits. Continuous operation is essential to successful mining; any production losses involve considerable monetary consequences. With ever-increasing productivity targets, there is a growing interest in mining automation. In line with expectations on digitization in industry in general, mining operations hope to experience a considerable paradigm shift through increased connectivity in the physical mines. Reliable wireless communication would enable both advanced data analytics and the advent of remote controlled, or even autonomous, heavy equipment such as LHD loaders (load, haul, dump machines).

Still, safety is the primary concern of any mining operation; improving worker safety trumps reaching increased productivity goals. However, increased mining automation also has the potential to increase worker safety by reducing the number of people working underground – although human workers will always need to enter the underground mine occasionally, e.g., to complete maintenance tasks and to inspect ongoing operations. Furthermore, the notion of safety-critical systems includes not only death or injury to people, also loss or severe damage to equipment as well as environmental harm is covered – and future underground mines will certainly rely on the operations of expensive machinery.

One of the most important aspects of safety in a mine is related to cave-ins. The structure of a mine is constantly changing, i.e., the extent of the crosscuts keeps changing as the ore body is processed. As the mining operations proceed, shafts and drifts are inevitably affected in position (inclination, rotation, lateral movements, and curves) and form (compression and deformation). Further factors that increase the risk of cave-ins include subsidence (gradual sinking of land), unsecured underground walls and ceilings, and cracks resulting from excessive excavation.

Regular shaft inspections are mandated by mine safety regulations, i.e., the determination of spatial changes of shaft columns that could be indicative of cave-ins. However, shaft inspections traditionally are time-consuming, and require halting of the mining operations. Furthermore, while the shaft inspections are critical, there is also a need to survey drifts and crosscuts. Laser scanning of mine shafts is already an area for which several companies offer services, but measuring horizontal structures has received considerably less attention.

The PIMM project, Pilot for Industrial Mobile communication in Mining [21], will push the limits for state of the art regarding new mobile communication technologies in heavy industrial mining environments. The project will result in business models for involved stakeholders in installation and operation of the infrastructure, as well as gains in productivity, maintenance and mining safety. Several use cases will be demonstrated, among them automation, supervision and remote control of heavy machinery.

As a byproduct of future autonomous vehicles operating in the Kankberg mine, there will be continuous LIDAR measurements of the mine drifts and crosscuts. In this report, we discuss the potential to use these LiDAR data to maintain a detailed 3D map of the Kankberg mine. Ideally, such a mapping system could operate without disturbing normal mining operations, and provide online availability of continuously updated geospatial information. Superimposition of subsequently scanned 3D images can then be used to determine and highlight issues such as ground movements, cracks in lining, and misalignment, analogous to what is currently the state-of-the-art offer for shaft inspections.

To investigate software business models and IT infrastructure in the future era of digitalized mines, we analyze combining LiDAR-equipped LHD loaders with Boliden's current mapping and positioning solutions into a system-of-systems (SoS). SoS is defined by Kotov as "large-scale concurrent and distributed systems of which the components are complex systems themselves" [13]. In a SoS, the constituent systems accomplish their own goals (i.e, by composing their respective elements into a system [2]), but the SoS functionality is more than the sum of its constituent systems. A key characteristic of a SoS is the independence of its constituent systems. Typically, they were developed independently, but through evolutionary development they are organized in a SoS, allowing novel goals to be met through emergent behavior [16]. Note that while we consider the combination of LiDAR and mapping solutions as a promising idea, it is mainly a means to stimulate SoS discussions in this report. Reliable wireless communication in Kankberg could enable also several other systems to be organized into a SoS – all in the same digitalization vein.

In this report, we analyze four alternative scenarios that illustrate different business models that could enable the SoS: 1) an in-house solution, 2) buying mapping capability as an add-on feature from the vehicle suppliers, 3) paying for mapping as an additional service from the hauling service provider, and 4) a software ecosystem. For each scenario, we present a high-level SoS architecture and discuss eight key variation points covering both technical and business perspectives. The amount of effort required by Boliden is discussed, clearly distinguishing between development and operations. Moreover, we do a SWOT analysis for each scenario to highlight our primary concerns. Finally, we conclude the report with a side-by-side comparison of the four scenarios based on the key variation points complemented by an outlook into what is needed next to realize the SoS.

The rest of the report is organized as follows: Section 2 presents the fundamental background, describing the Boliden context and the Kankberg mine. The same section introduces mapping from the perspective of mining safety, the PIMM project, and essentials of LiDAR for autonomous LHD loaders. Section 3 presents related work on shaft monitoring and robotic mine mapping, both areas related to the SoS we propose. In Section 4, we introduce the fundamentals of the proposed SoS, and describe the assumptions we make for all four scenarios. The section emphasizes the SoS architecture and the concept of architecturally significant requirements. Section 5 is the main contribution of the report, presenting the four SoS scenarios. In Section 6, we discuss some implications of our results. Finally, Section 7 concludes our findings and suggests what should be the next step toward the SoS.

## 2 Background

This section covers Boliden background information and a description of the pilot mine Kankberg. We also introduce mining safety and some ongoing work on remote controlled and autonomous LHVs.

### 2.1 Boliden and the Kankberg Mine

Boliden is a leading metals company with core competency within exploration, mining, smelting, and metal recycling. The company has approximately 5,500 employees. The driving market forces for metals companies are increasing demand due to trends in infrastructure, the construction industry, and the automotive market. More specifically, a recent Boliden report identifies population growth, urbanization, and economic growth in developing countries as key business opportunities. Boliden's overall strategy involves stable and efficient operations, organic growth, and selective acquisitions.

Boliden's success in the metals business relies on state-of-the-art technology, including efficient mine design, mobile control systems, and increased automation; the latter being a strong trend in mining operations globally [5][9]. Boliden has an explicit ambition to run projects in-house, i.e., internal know-how is considered fundamental. The current automation trend promises increased productivity through the introduction of autonomous machines, wireless data transfer, and positioning of people and equipment – all these solutions rely on software-intensive systems. However, whether Boliden will be able to keep all required technical software expertise in-house is uncertain. The scaling role of software in traditional industries has attracted considerable research efforts lately [8], and turning into a software-intensive company is an acknowledged challenge. The IT department at Boliden roughly employs 100 people including support functions – only a fraction of them deal with systems resembling the SoS envisioned in this report.

Boliden utilizes six mining areas in Sweden, Finland, and on Ireland (cf. Figure 1). The Boliden Area is located in the mineral-rich Skellefte field in the north of Sweden, which has been mined by Boliden since the 1920s from more than 30 different mines. Today, 500 employees work in the area consisting of a concentrator at Boliden, and four nearby mines. Three mines are underground: Kankberg (gold and tellurium), Renström and Kristineberg (polymetallic, i.e., zinc, copper, lead, gold, and silver), and Maurliden is an open-pit mine awaiting reclamation (polymetallic). In total, approximately 1,700 ktonnes of ore are mined and concentrated in the Boliden area every year. The copper and gold concentrate produced is sent to both Boliden smelters in Sweden, Norway, and Finland, but concentrate is also shipped to external smelters across Europe.

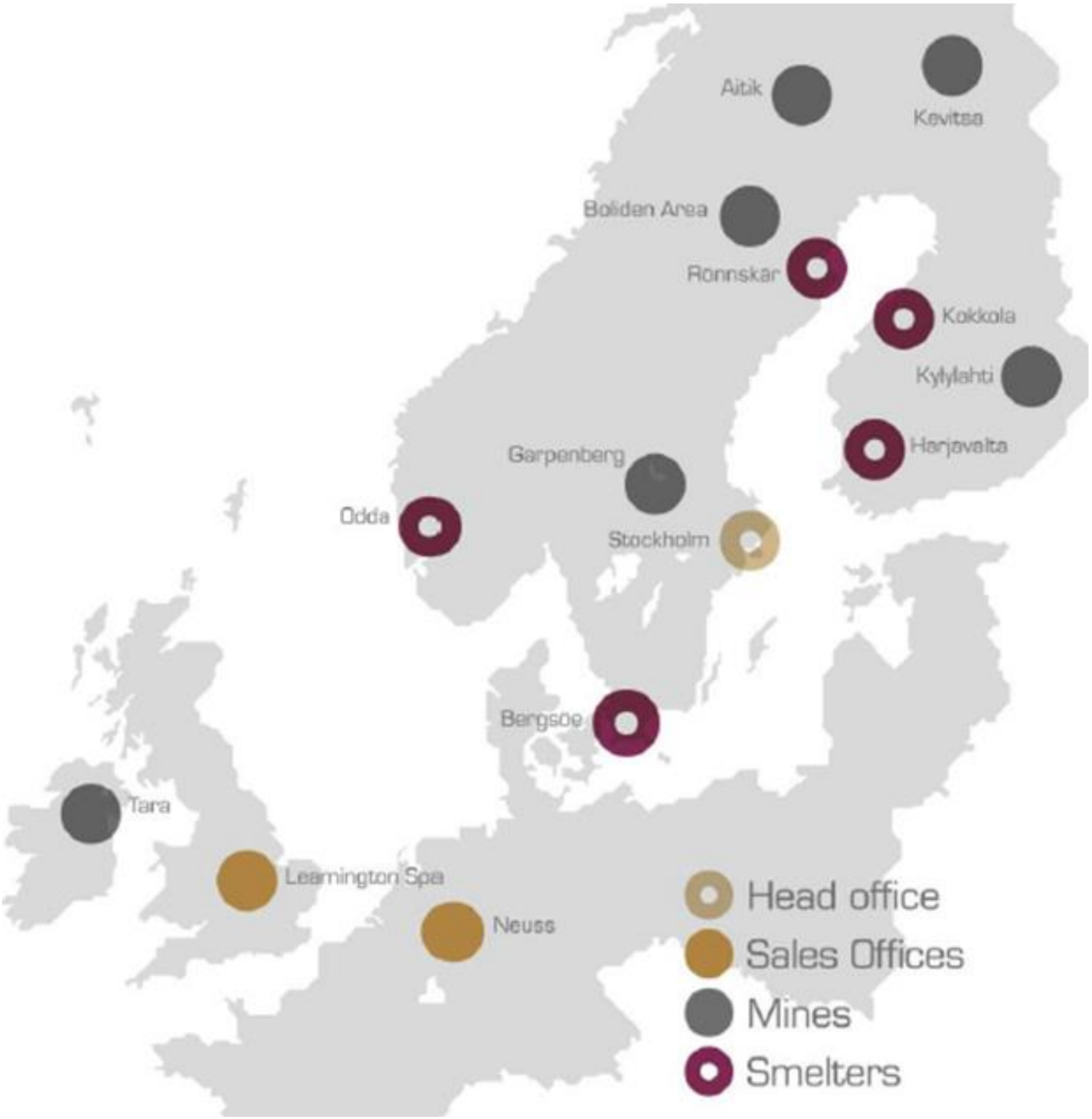

Figure 1 Boliden sites across Europe. The Kankberg mine is in the Boliden Area in the North of Sweden. [Illustration: Boliden Press]

The Kankberg mine was closed in the 1990s, but it was reopened in 2012 due to the discovery of a new gold and tellurium ore body - 250 kg of gold was produced already during the first year. After major investments, Kankberg is today 500 m deep. In the mine, the cut-and-fill method is used with granite as the backfill material. Drilling and blasting is used to break rock for excavation, i.e., controlled use of explosives. The mine employs about 80 people and around 20 contractors. Kankberg has an expected life of mine until 2020. Figure 2 shows an overview of the Kankberg mine.

In mining terminology, *shafts* are vertical tunnels used for ventilation and transportation. Horizontal tunnels are called *drifts* (cf. the main access drift in Figure 1). Drifts across the ore body are referred to as *crosscuts*.

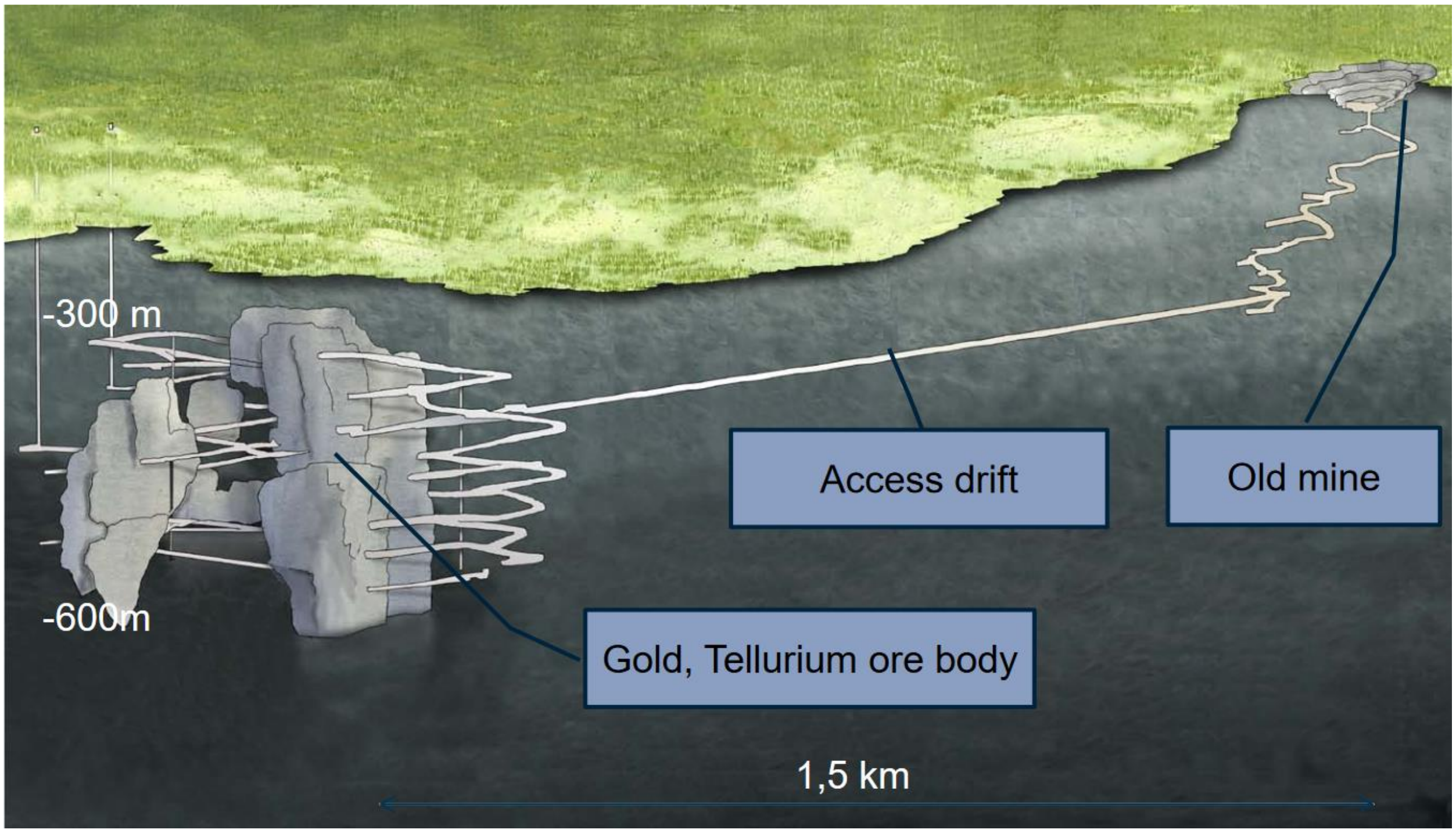

Figure 2: The spatial structure of the Kankberg mine. [Illustration by Boliden Press]

### 2.2 Mining Safety and the Importance of Mine Mapping

Safety is the top priority in any mining operation. Boliden has a zero vision for accidents in 2018, and is determined to use the best available technology and processes to reach the goal. Several risks associated with working in an underground mine must be considered. Potential accidents include cave-ins, flooding, gas explosions, chemical leakage, and electrocution. Furthermore, there are long-term health risks related to mineral dust, radon, welding fumes, mercury, noise, and heavy loads. The potential SoS discussed in this report focuses on threats related to cave-ins.

There are three main causes of cave-ins in underground mines. First, hasty mining operations might fail to secure walls and ceilings of shafts and drifts. Second, excessive excavation might lead to cracks in floors and walls, thus weakening the entire structure. Examples include insufficient vertical spacing between crosscuts and too rectangular crosscuts causing stress concentration in corners. Third, gradual sinking of land can cause cave-ins [24], i.e., *subsidence* (the downward motion of a surface). As illustrated in Fig. 3, mining operations induce subsidence of the Earth’s surface. While mining-induced subsidence is rather predictable in magnitude and extent, monitoring the progress is fundamental to mining safety. In Kankberg, however, the mountain stresses caused by drilling and blasting activities dominate any subsidence.

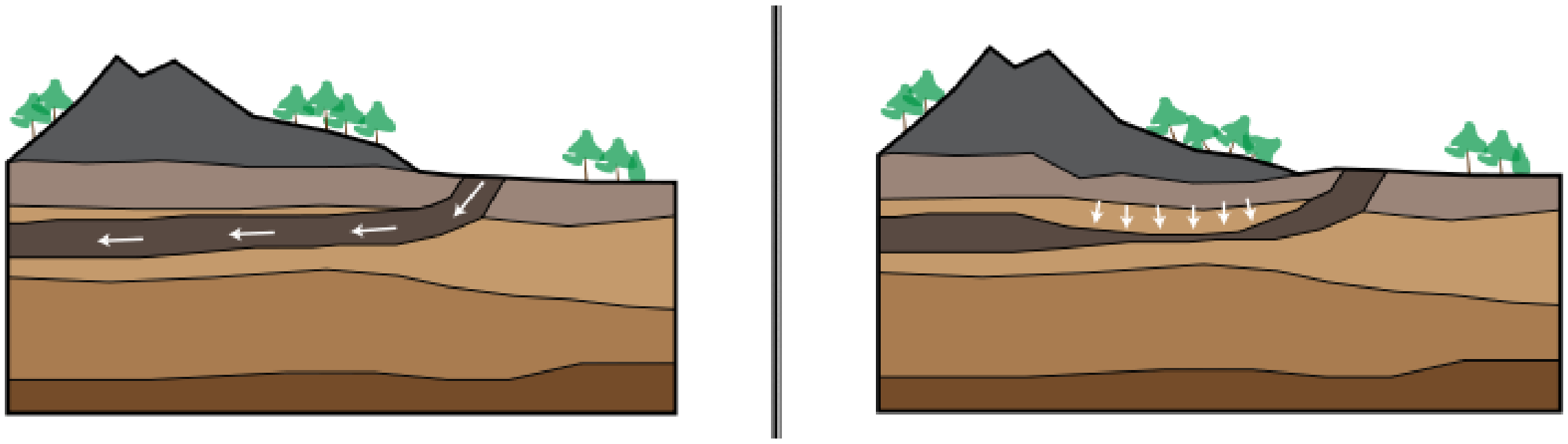

Fig 2: Subsidence caused by underground mining. Illustration by Mpetty1[1].

In Kankberg, Boliden's primary equipment to perform stress-strain measurements are extensometers and cable bolts. In areas directly impacted by drilling and blasting, optical measurements are conducted, i.e., manually collected data comparing distances to reference points with a millimeter precision. One promising approach to monitor changes in rock mass plasticity, originating in either drilling and blasting or subsidence, is to maintain highly accurate maps of the underground mine. However, for the maps to be truly useful from a safety perspective, such maps need to be as precise as current optical measurements. Another reason to maintain accurate maps is commercial. The creation of *volumetric* maps is critical to the mining business, i.e., maps that estimate the value of an ore body.

### 2.3 Remote controlled LHDs in an Underground Environment

Improved productivity and efficiency in production is the key for continued leadership in Swedish industry in all production disciplines. The ongoing digital revolution and the ability to incorporate new mobile communication into production systems will further improve competitive advantage for Swedish industry. The PIMM project, Pilot for Industrial Mobile communication in Mining [21], will push the limits for state of the art in the combination of new mobile communication technologies and heavy industrial mining environment. PIMM will also result in business models for involved stakeholders in installation and operation of the infrastructure, discussing potential productivity gains, reduced maintenance costs, and increased safety. Several use cases will be demonstrated, among them automation, supervision and remote control of heavy machinery in the Kankberg mine. The partners collaborating in PIMM are: Boliden, Ericsson, ABB, RISE SICS, Volvo Construction Equipment, Wolfit, Luleå University of Technology, and Telia Sonera.

Project goals:

- To evaluate new mobile communication infrastructure in an industrial context where the PIMM partners can validate the technology, applications and business models.
- The project will work with the strict requirements for safety, robustness and productivity required of operations in underground mines.

[1] Own work, CC BY-SA 3.0 - https://commons.wikimedia.org/w/index.php?curid=14698311

To demonstrate the usefulness of a connected mine, a number of use cases will be realized:

- Remote controlled machinery, where the driver is at the surface of the mine and the LHD is remotely driven, using the 5G network in the mine.
- The use of smartphones in the mine, i.e., in locations where you normally do not have any coverage.
- Wireless sensors for temperature, humidity, etc.
- Connected automation, e.g., enabling the control of ventilation over a wireless network.

For Boliden, the long-term goals enabled by reliable wireless communication are:

- Improved safety
    - Fewer people under ground
    - More connected equipment
    - Robust communication
    - Seamless integration with rescue personnel at accidents
- Improved productivity
    - Increased automation of mining operations
    - Seamless integration with open air mobile communication
    - Supervision of equipment and predictive maintenance

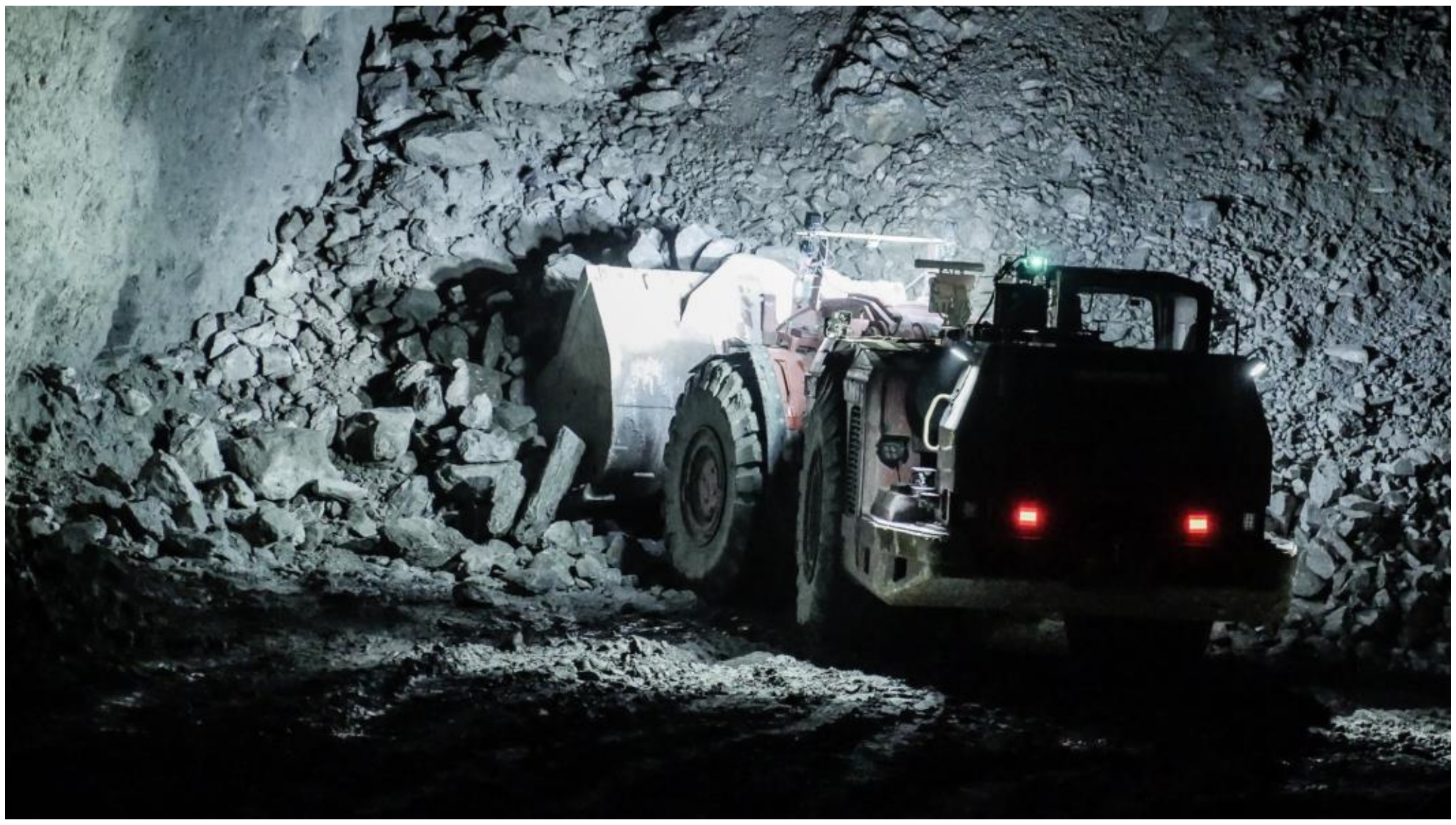

Fig 3. The remote controlled LHD.

LHDs operating underground will rely on LiDAR for navigation. LiDAR (Light Detection and Ranging) is a measurement system consisting of three components: lasers, GPS (Global Positioning System) and IMU (Inertial Measurement Unit). The latter is, in essence, an optical gyroscope. The integration of the three components allows high accuracy measurements of the coordinates of the point in space where a laser beam reflects from a non-transparent object. LiDAR is used for various applications, from construction industry and road maintenance to city management and planning [23].

Mine environments present certain challenges for laser scanning systems [12]. First, there is a risk that open beam lasers could ignite methane gas. The US National Institute for Occupational Safety and Health (NIOSH) report that laser beams should not exceed 150 mW or 20 $mW/mm^2$. Second, underground mines present areas with widely-ranging albedo, e.g., white and black patches. Third, shiny metal and wet surfaces in underground mines challenge the laser scans by introducing noise. All these factors must be considered when mounting LiDAR on autonomous LHD haulers.

## 3 Related Work

This section presents commercial offers of spatial measurements in mines, as well as state-of-the-art mine mapping using robots from the research front.

### 3.1 Commercial Solutions for Shaft Surveying

SightPower offers a LiDAR solution for shaft surveying that creates a 3D scan of the current shaft conditions. Subsequently, the shaft scanning system can be used as frequently as required, e.g., on a weekly basis. By comparing the subsequent scans, it is possible to analyze all changes and their severity. LiDAR measurements outperforms the accuracy of manual inspections relying on the human eye, and removes issues related to human fatigue, loss of concentration, and poor lighting conditions. Thus, the improved measurements enable earlier detection of defects in shaft infrastructure, i.e., increasing the safety of mining operations. SightPower combines the LiDAR measurements with a proprietary system for recognition of constructive elements. The system supports planning of maintenance activities, by detecting changes of shape and location, as well as level of corrosion.

DMT offers a solution referred to as Inertial Shaft Survey for Mining (ISSM), a measurement system based on three accelerometers and three gyroscopes (rotational rate sensors). The unit is mounted on a carrier, and every change in its orientation produces a gyroscope signal that is used to determine the precise position of the carrier. By correcting the amount of rotation to the natural rotation of Earth, the integration of the accelerometer data provides the position of the carrier. The accuracy of the measurement system is 50 mm per 1000 m horizontally, and 100 mm per 1000 m vertically, respectively. However, the ISSM system is limited to measuring the spatial position of the carrier, i.e., the cage on which the carrier is mounted. Also, the ISSM system is exclusively designed for measuring mine shafts. The system we are suggesting on the other hand, delivers highly accurate measurements of the mine drifts.

### 3.2 Robotic Mine Mapping

Several researchers have worked on mine mapping using autonomous robots. The research has mainly targeted abandoned mines, a substantial problem in the US; not even the US Bureau of Mines are certain of how many abandoned mines exist in the country. Accurate maps of the old mines are typically missing, but could be of great value to avoid catastrophic events [20]. The motivation for fully autonomous robots has been twofold. First, abandoned mines present harsh environments, dangerous to humans. Dangers include lack of structural soundness, low oxygen levels, and risk of flooding. Second, the inadequacy of current wireless communication techniques makes remote controlled robots unfeasible.

Thrun and colleagues have conducted several experiments on autonomous mapping of abandoned mines. Their robot Groundhog evolved into a 1,500-pound robotic vehicle, equipped with, e.g., onboard computing, laser range sensing, gas sensor, and a video recorder [24]. They have reported a series of successful mapping operations [3]. However, limited by the technology at the time, the robot did only perform 3D scans at regular vantage points; using a tilting mechanism to acquire a point cloud of the area in front of the robot. Current autonomous vehicles acquire 3D point clouds more frequently.

Huber and Vandapel and their research group also did work on underground mine mapping using robots [12]. In contrast to the work by Thrun et al., Huber and Vandapel tried their approach in active mines. They mounted a high-resolution 3D scanner on a mobile robot, providing 8000 x 1400 pixel scans with millimeter-level accuracy. As for Groundhog, they collected scans only at certain vantage points; Each three to five meters the robot stopped for 90 seconds to obtain a complete scan. A considerable contribution of their research is related to multi-view surface matching, i.e., merging multiple 3D views into a single map. Their approach is called iterative merging, which was successfully used to create high-quality maps of an underground mine. However, their approach does not scale to large numbers of scans; Back in 2006, their approach could only generate sub-maps containing about 50 scans.

## 4 Technical Background: Introduction to a Potential Solution

In this section, we introduce the preconditions and general issues that the envisioned SoS must address. Fig. 4 lists all assets that are involved in the SoS, organized into the three dimensions *acquire*, *adapt*, and *(re)use* [19]. The assets are of the following types: software systems, software components, cyber-physical systems, and IT infrastructure. Assets are organized along an y-axis depicting the level of new development required. Constituent software systems are presented as dark gray boxes while software components are white. Also, existing LHDs that need to be adapted are presented, as well as IT infrastructure that either is reused or acquired.

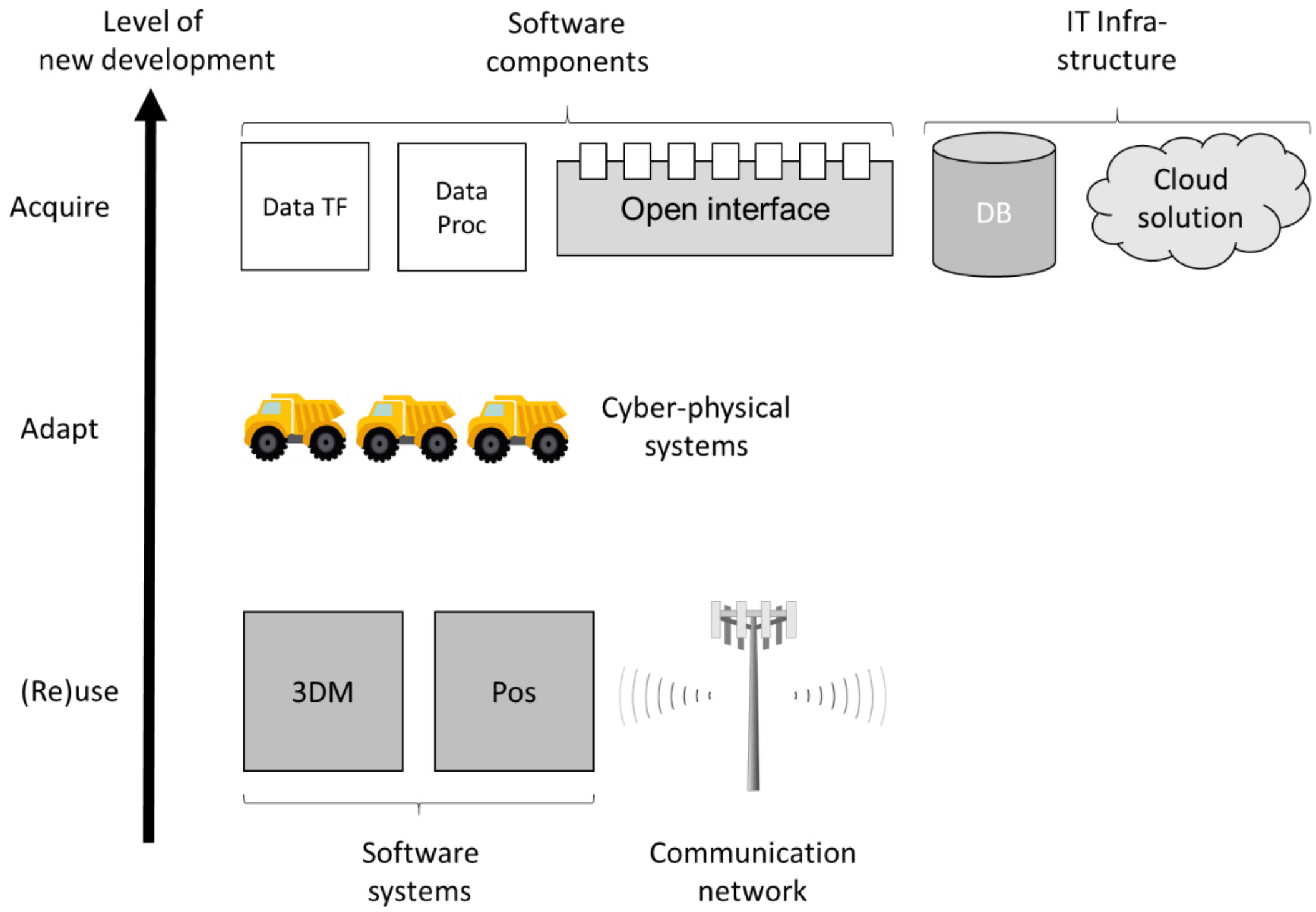

Figure 4: Overview of the SoS parts involved in the solution proposals.

We assume, in our scenario, the availability of a reliable high-bandwidth wireless communication network in the Kankberg mine, e.g., a 5G public mobile network as piloted in the PIMM project [21]. Furthermore, we assume that LiDAR equipped LHDs are operating in the mine and that necessary adaptation to make the continuous stream of LiDAR accessible will be done. Furthermore, we base our designs on two existing systems already in use:

- 3D Map rendering software (3DM), a commercial off-the-shelf 3D CAD solution used by the mining engineers. Today, there are no 3D maps being rendered on a regular basis, only contour maps are available for Kankberg. We use "3DM" to refer to the mapping software in general terms, examples of specific solutions include MicroStation, Deswik.CAD, Datamine, and Leapfrog.

- Pos, a positioning system for vehicles, equipment and personnel. Already today, Boliden has a software for Pos, which monthly imports map updates from 3DM.

The main challenge in designing and realizing the SoS is to organize the responsibilities of Boliden and their partners, incl. software vendors, vehicle suppliers, and service providers. In this work, we distinguish between development (Dev) and operations (Ops) [4] – and discuss pros and cons of different constellations and their corresponding business models.

To structure our discussion, we present four contrasting SoS scenarios from Boliden's perspective:

1. *Scenario In-house* – Boliden is system integrator and coordinator of IT subcontractors.
2. *Scenario* A*dd-on option* - vehicle suppliers sell the mapping feature as an add-on option.
3. *Scenario Additional service* - an external hauling company sells mapping as an optional service.
4. *Scenario Ecosystem* - a semi-open ecosystem enables a third-party mapping solution.

Each individual scenario can be realized in numerous ways, but for the clarity of this report, we present one complete possible design for each case. On the other hand, certain aspects of the SoS design will be the same across all four scenarios. We identify these by exploring the seven key questions stated by the strategic SoS research and innovation agenda for Sweden [2]. We argue that the answers to the first five questions are the same:

***Why* is it created*?*** The new SoS is created to improve safety and productivity in the underground mine. The SoS should utilize the fact that LiDAR equipped LHDs are operating in the mine, and reuse the obtained 3D point clouds to update 3DM and in turn Pos.

***Whose* system is it?** Boliden takes ownership of the SoS. Agreements must be made with the vehicle suppliers to acquire 3D point clouds from the LiDAR equipped LHDs.

***Who* are the stakeholders?** Boliden is the primary beneficiary of the new SoS. Secondary beneficiaries include stakeholders of the constituent systems: vehicle suppliers that can improve their predictive maintenance, and Pos, whose product offer benefits from pinpointing of equipment and personnel on exact maps. Furthermore, safer working conditions are in the interest of the labor unions.

***What* should it do?** The SoS should enable continuous updates of Kankberg maps at unprecedented level of detail.

***How much* should it perform?** The main quality attributes of the SoS is performance and mapping accuracy. The mapping accuracy should be on a sub-centimeter scale, and the computational performance should allow map updates at least on an hourly rate.

Finally, the remaining two questions will be addressed differently across the four scenarios. The main contribution of this paper is SWOT analyses for the four scenarios, in the light of the following two questions:

> ***How* should it be organized?** Should the SoS architecture be organized as an in-house, add-on option, or additional service solution? Or would a semi-open ecosystem be feasible? How can interoperability be achieved? What mechanisms are needed for managing the SoS operations?
>
> ***When* does it change?** What are the potential evolution scenarios for the SoS?

### 4.2 Architecturally Significant Requirements

Architecturally Significant Requirements (ASRs) are the requirements that are more influential on the system architecture than others. From Clements and Bass [7]:

> *Most of what is in a requirements specification does not determine or — "shape" an architecture. Architectures are mostly driven or shaped by quality attribute requirements. These determine and constrain the most important architectural decisions. And yet the vast bulk of most requirements specifications is focused on the required features and functionality of a system, which shape the architecture the least. Worse, most do a poor job of specifying quality attributes; many ignore them altogether.*

Besides having a larger impact on the architecture, they tend to be vague and context dependent [6]. As pointed out by Clements and Bass [7] and Anish *et al.* [1], non-functional requirements (a.k.a. NFR or quality requirements). The NFRs we consider in this report are:

- Security
  Security includes integrity of data and users, access control (including identification and access logs, etc.) and transmission of data in a secure manner.
- Reliability
  Reliability is sometimes called dependability. Typically, availability of a service, loss of data and time to recover from a failure are included.
- Efficiency
  The efficiency requirements are of the types response-time, use of resources (such as storage or battery), bandwidth, etc. Efficiency requirements are often called performance requirements.

## 5 Four Alternative Realization Scenarios

The envisioned SoS could be realized in several different ways. In this section, we present the four contrasting scenarios introduced in Section 4: 1) in-house, 2) add-on feature, 3) additional service, and 4) ecosystem. For each scenario, we present a high-level figure illustrating both the technical perspective and the business perspective.

To allow a coherent discussion, we highlight the same list of key variation points for each scenario, again organized into a technical and business scenario. Table 1 describes the 8 variation points discussed across the scenarios. Finally, we conclude each scenario with a SWOT analysis (strengths, weaknesses, opportunities, and threats, a structured planning method [22]).

For all four scenarios, we make the following technical assumptions:

- Boliden owns or leases the vehicles but not necessarily all services around them.
- Reliable high-bandwidth data communication is available in the Kankberg mine [21].
- LiDAR equipped LHDs from different vehicle suppliers operate in the underground mine.
- LiDAR data can be exported from the LHDs, at least from some vehicle suppliers, not necessarily without cost though or adaptation.
- 3DM can import data from external sources.
- The format of the LiDAR data, i.e., detailed 3D point clouds, is not standardized. However, all formats can be transformed into a valid 3DM import format.
- Pos can import data from 3DM and this operation is not a heavy one.
- Sufficient computational resources are available to perform all needed data processing, e.g., balanced using the elasticity of cloud computing [10].
- Unrestricted data storage is available, either from an internal data center or from an external cloud storage provider [25].

Table 1: Key variation points discussed for each scenario.

| | Name | Description |
|---|---|---|
| Technical perspective | IT Infrastructure | The IT infrastructure consists of databases, servers, network, etc. An important aspect is also whether components are in the cloud or use cloud resources. Requirements on the network communication depends heavily on the infrastructure. Other aspects include where storage, which will need to store a significant amount of data, is placed and whether heavy computation is performed. |
| | Evolution | All IT system evolve over time, including the IT systems already in place at Boliden. The introduction of the mapping SoS is a type of evolution. Evolution in this report refers to the decision whether to make (develop) internally or buy, how scalable the system is to changes in the operational environment and how much control Boliden has on the roadmaps of the constituent systems. |
| | ASRs | Architecturally Significant Requirements (ASRs) are requirements that have a considerable impact on the IT infrastructure or the architecture. In this report, we focus on security, reliability, and efficiency. |
| Business perspective | Inter-company dependency | How much does the SoS design depend on a fixed set of collaborating companies? To what extent does vendor lock-in occur? We analyze how proprietary the solutions in the scenarios will be and what this means to the business relationships. |
| | Upfront investment | Upfront investment is coupled with the decision on whether to make or to buy solutions. Under this key variation point we discuss fixed initial costs that involved in developing the SoS. |
| | Running costs | Running costs instead cover expenses from subscription fees or operations costs. As opposed to upfront investments, these costs are paid on a regular basis as the SoS is used. |
| | Risk | There are many risks to consider when develop a SoS. We focus on risks associated with unsatisfactory operations of the SoS, i.e., an inadequate mapping solution. Also, we consider risks regarding sensitive information falling into the wrong hands. |
| | Innovation platform | Innovation is coupled to company culture and the business model. Boliden is not in the mapping business, hence unlikely to be innovative in this area – ownership of a mapping solution is unlikely. Thus, Boliden is dependent on the partners in the business network for innovation. Depending on the setup of the business model and the companies that can be included, the innovation platform can be more or less strong. |

### 5.1 Scenario In-house - Boliden is System Integrator and Coordinator of IT Subcontractors

In the first scenario, Boliden takes on the leading role in designing the SoS and acts as owner of the system integration. While an IT subcontractor does a large part of the actual development effort, Boliden undertakes initial requirements elicitation and contribute actively in the iterative specification of requirements.

Figure 5 outlines the scenario. In the upper part of the figure, Boliden pulls the LiDAR data from the LHDs supplied by VS1 and VS2 through open APIs (cf. A). In this scenario, LHDs from VS3 do not support access to any LiDAR data. Both LiDAR data from VS1 and VS2 are transformed, using two customized software components (cf. B), before being stored in the Boliden data storage (cf. C). The LiDAR data is then used in a data processing component (cf. D), filtering out measurements that differ from previous point clouds. Moreover, the data is pre-processed into valid 3DM import format. Finally, the LiDAR data is imported by 3DM to update its 3D CAD models (cf. E) – repeatedly synchronized with the maps in Pos (cf. F).

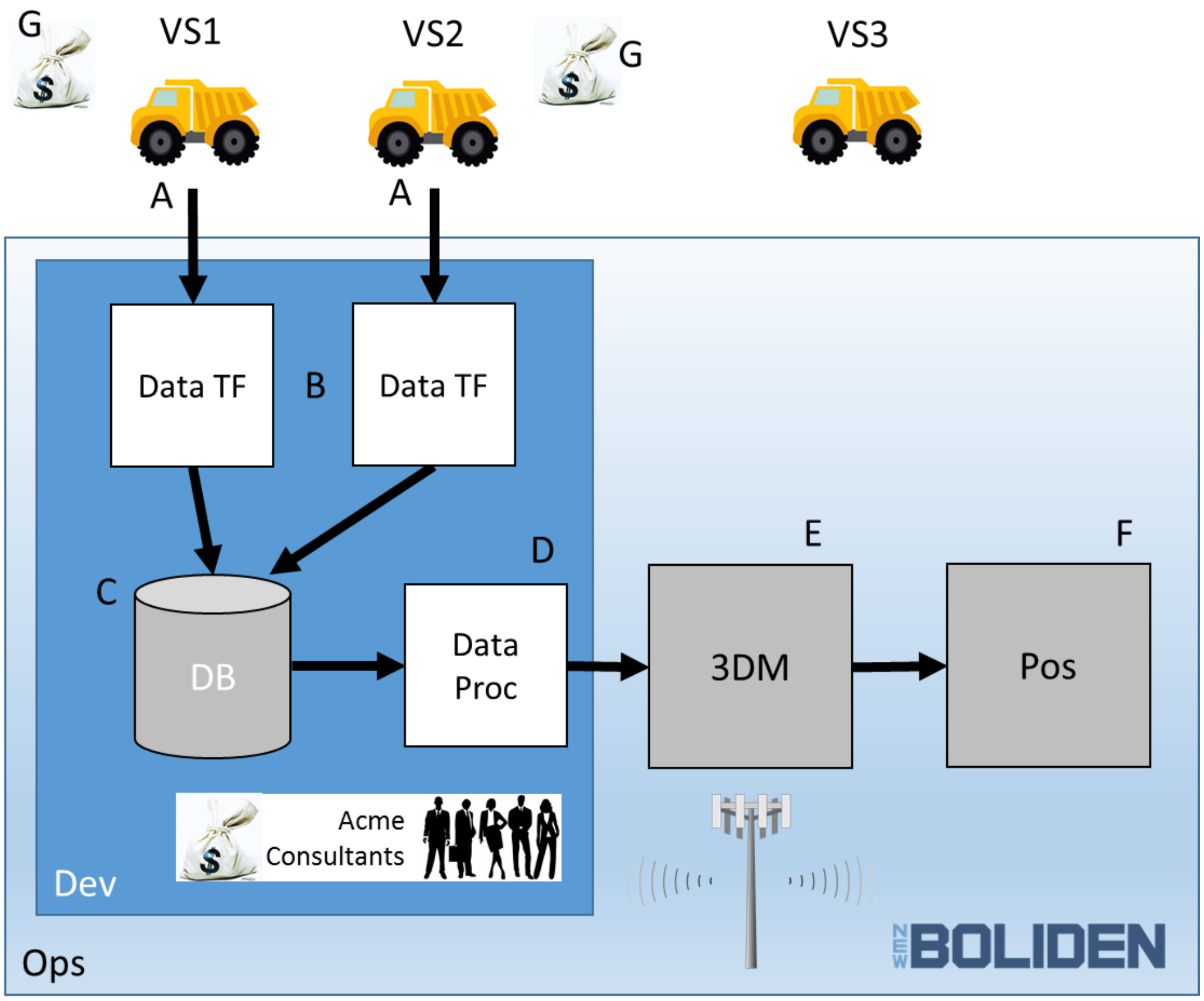

Figure 5. Overview of Scenario In-house. Boliden owns both the system integration and the operations.

Figure 5 also describes the scenario from a business perspective. Boliden acquires development resources from Acme Consultants, which means external development of the data transformations, the database solution, and the data processing (cf. area Dev) for an upfront investment. Boliden signs an agreement with VS1 and VS2 to allow extraction of the LiDAR data, realized through upfront investments (cf. G) but no running costs and no technical support. Apart from this agreement, the vehicle suppliers are not involved and do not obtain any direct benefits from the SoS. From an operations point of view, Boliden is solely responsible for the entire SoS, i.e., both the constituent systems and the new software components. Evolution of the SoS, including adapting to data and API changes of the LHDs, are not straightforward in this scenario as it requires additional resources acquired from Acme Consultants.

Regarding the key variation points, we describe Scenario In-house as follows:

*Technical perspective*

#### IT infrastructure

In Scenario In-house, all constituent systems are hosted by Boliden and using local resources. The systems are connected to the local high-bandwidth communication network and all data are kept on Boliden's local data storage. The LiDAR data are typically pulled from the vehicles by a system that also transforms the data into a common format. Also computation is performed locally. Most of the implementation is on systems under Boliden's control and will be tailored to the specific needs of Boliden.

#### Evolution

As the name of Scenario In-house indicates, development is done by Boliden, through a contractor though. Hence, the resulting SoS will be proprietary to Boliden. As Boliden owns both the infrastructure and the mapping SoS, how easy it is to adapt the SoS for changes in the environment is completely up to Boliden. However, as development and operations is handled locally, scalability is likely an issue. On the one hand, there is little competence and resources in-house meaning there is a need to source in a contractor, on the other hardware is local and to scale up means buying more computers. Furthermore, as the solution is tailored to the specific needs of Boliden and the specific suppliers, etc., there is a risk that introducing a new supplier e.g. for the LHD, or new features will mean development effort is needed to integrate them into the SoS. Upgrades of constituent systems also poses a risk of being cumbersome. Switching to another solution is also likely to be non-trivial, e.g. another map solution or hardware platform. As the systems are under the control of Boliden and any development effort is paid by them, they have complete control of what is being developed as well as the release schedule. They do not need to adapt to available solutions or compromise with the functionality of other systems.

#### ASRs

*Security*: All computation and rendering of maps and usage of the data is performed by Boliden. No other company has access to any information, i.e., security is not a big issue.

*Efficiency*: In terms of scaling of the computation, Boliden is largely left on their own. Hence, efficiency requirements to ensure a scalable solution are important early on to avoid expensive refactoring late in the project or the need to add hardware later in the operations phase which might not even be possible without updating the software.

*Reliability*: Reliability is unlikely to be a driving requirement. While availability of course is important, the communication infrastructure is simpler in Scenario In-house compared to the others. Furthermore, as Boliden operates all IT systems, their internal routines and procedures outside the actual software implementation should handle any availability problems.

#### *Business perspective*

##### Inter-company dependency

Boliden customizes data access from LHDs supplied by VS1 and VS2. Moreover, the data processing is tailored to support the input format of 3DM. Consequently, the relationship deepens with selected companies, making Boliden more reliant on them and making it more difficult to switch both vehicle suppliers and mapping solution. Hence, relationships tend to be static and stable.

##### Upfront investment

Boliden invests in internal solutions for data storage and computation as well as the development effort for the mapping SoS. Also, the SoS integration is outsourced to Acme Consultants as a complete development project.

##### Running costs

Boliden recruits engineers to maintain the solution once deployed, e.g., to cover maintenance of the data center hardware. Some resources from Acme Consultants are also acquired to meet the needs for software maintenance of the SoS.

##### Risk

As Boliden is responsible for operations. Hence, Boliden must themselves solve problems that arise. Hence, Boliden carries all risk and must ensure mitigation internally. This might mean spending effort on monitoring and issue resolution if something occurs. Also, should there be a hardware problem, there is a risk of long down-time.

Boliden owns all collected LiDAR data, and the data are stored internally. While the LHD suppliers have access to the LiDAR data from their respective vehicles, the Scenario In-house still provides the highest data privacy among the four alternative scenarios.

##### Innovation platform

Development of this type of system is not the core business of Boliden and the available resources are limited. As a result, this setup does not promote further innovation neither within Boliden nor by the vehicle suppliers. We assume that Acme is just suppling development resources, hence is unlikely to take an interest in driving innovation.

We conclude the description of Scenario In-house with the SWOT analysis below.

Table 2. Scenario In-house SWOT analysis

<table>
<tr><td>Strengths<br>• Data privacy<br>• Roadmapping</td><td>Opportunities<br>• Could be fast to implement, if Boliden is ready to invest</td></tr>
<tr><td>Weaknesses<br>• All R&D costs and risks<br>• Missing competence in-house<br>• Proprietary solution → no sharing of costs with others</td><td>Threats<br>• Expensive evolution<br>• Technical debt.<br>• Suppliers see business opportunity → future charges difficult to foresee<br>• Lack of knowledge internally</td></tr>
</table>

### 5.2 Scenario Add-on Option - Vehicle Suppliers Sell the Mapping Feature as an Add-on Option

In the second scenario, some vehicles suppliers have recognized mapping as a promising business opportunity; the positioning sensors are mounted on their LHDs after all, why not make a profit by selling the LiDAR data to Boliden? They decide to offer the mapping feature as an add-on option. In this scenario, we illustrate how LiDAR data might be sold to Boliden in different formats, i.e., either in a raw format or processed for direct import into 3DM. Scenario Add-on Option relies heavily on distributed resources.

Figure 6 presents an overview picture of the scenario. The top of the figure shows VS3's LHDs storing LiDAR data locally. Each individual VS3 LHD implements the entire pipeline (cf. A): data collection, data storage, data processing (incl. filtering), data transformation, and data transfer to Boliden's 3DM, i.e., both the data storage and computational resources are distributed. VS1 provides a similar add-on option to their LHDs (cf. B), but without the data transformation component. To transform the LiDAR data into the import format of 3DM, the SoS thus uses a customized software component developed by Acme Consultants (cf. C). In this scenario, VS2 does not offer the mapping feature as an add-on option. Finally, the LiDAR data is imported by 3DM to update its 3D CAD models (cf. D) – repeatedly synchronized with the maps in Pos (cf. E). Depending on the contracts with the suppliers, it is of course possible that Boliden could access the raw data in the databases, e.g. to use the data for other types of analysis. We do not elaborate on this here.

From a business perspective, Boliden initially buys the mapping feature as an add-on option when purchasing LHD's from VS3 (cf. F) – a feature that implements the whole chain from LiDAR measurements to import into 3DM, but it requires Boliden to pay a subscription fee. Boliden also buys a cheaper but less advanced add-on option from VS1 (cf. H), i.e., delivery of raw (but filtered) LiDAR data. To acquire the necessary data transformation component, Boliden outsources development to ACME Consultants for a fixed cost (cf. area Dev). Both add-on options bought from the vehicle suppliers comes with warranty and support, however limited to the specific mapping feature, i.e., all steps prior to 3DM import. There are no additional running costs from buying the add-on feature.

The vehicle suppliers do not obtain any direct benefits from the SoS itself, but in contrast to Scenario In-house, they are processing LiDAR data – suggesting that the vehicle suppliers employ data analytics, facilitating predictive and preventive maintenance of the LHDs [15], or at least collecting valuable operations data from the field. In Scenario Add-on, Boliden is no longer sole responsible of the SoS. Instead, the Boliden SoS operations is primarily focused on enabling uninterrupted data flow and transformations from VS1's mapping feature – the VS3 counterpart is an external concern. The evolvability of the Scenario Add-on Sos is limited. Any novel features originate in development on the side of the vehicle suppliers, thus only driven by their expected profit. Furthermore, Boliden needs to acquire development resources from Acme Consultants when VS1 makes changes to their LiDAR data format.

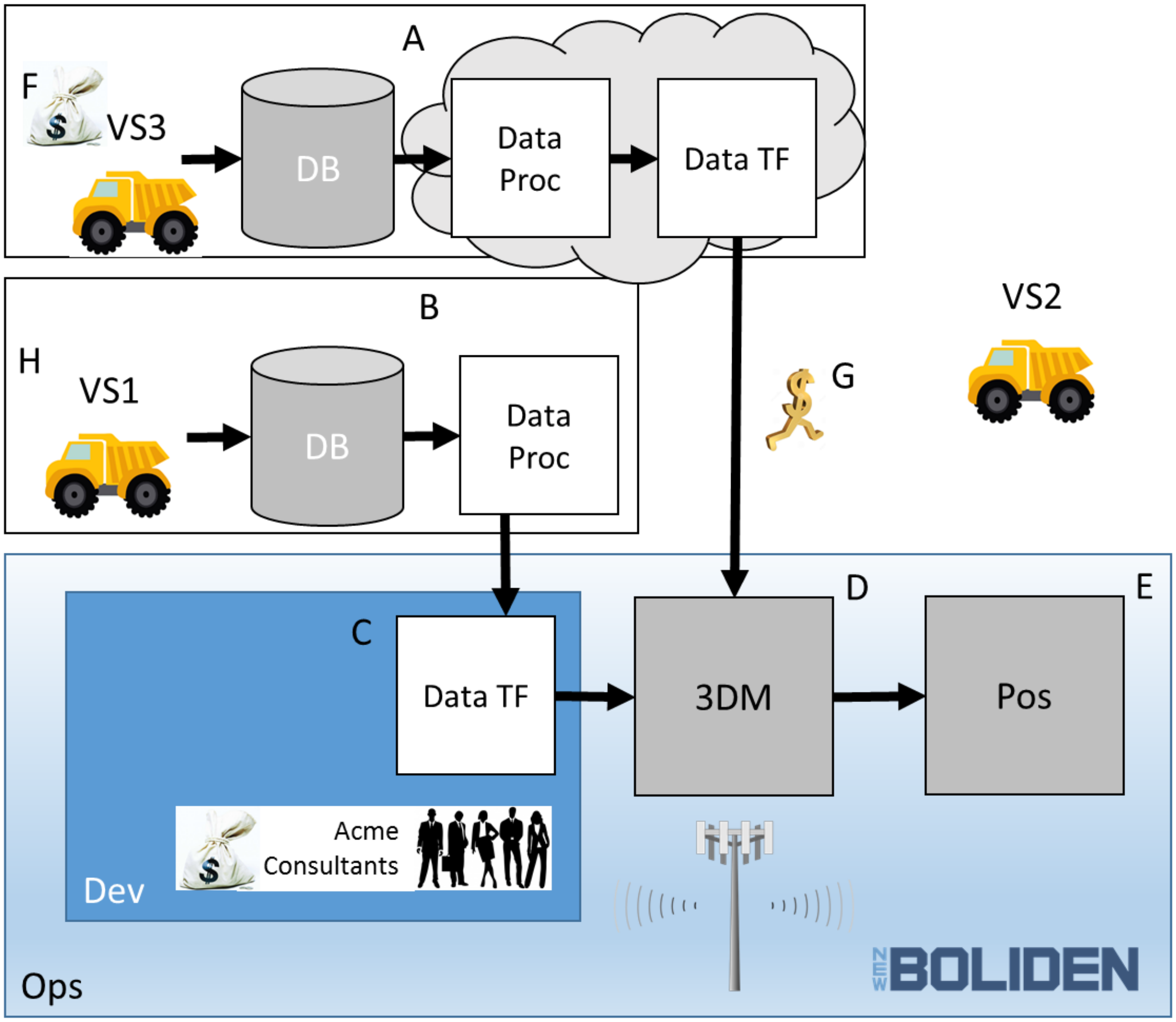

Figure 6: Overview of Scenario Add-on Option. Some vehicle suppliers sell mapping as a feature, with or without data transformation for a subscription fee.

Regarding the key variation points, we present Scenario Add-on Option as follows:

*Technical perspective*

### IT infrastructure

In this scenario, the vehicle suppliers have a larger role as to provide an add-on service to their vehicles. The suppliers would handle the extracting of the LiDAR data and the computation to render a map. VS3 uses cloud computing to process and transform LiDAR data before transferring it to Boliden. As there is no guarantee that the suppliers will use a common format nor that it works with the 3DM solution the data from the suppliers might need to be transformed; In Scenario Add-on Option, Boliden develops and hosts such a solution for data collected LHDs supplied by VS1.

#### Evolution

The vehicle suppliers offer a generic feature for the open market. Adaptation to specific needs at Boliden are not likely, instead Boliden needs to adapt to the vehicle suppliers. The Boliden part of the SoS is rather small and most of its evolution originates in the R&D of the vehicle suppliers. Adding additional Boliden features is difficult, such as novel data analytics. On the other hand, LiDAR data from additional vehicle suppliers could be incorporated in the SoS, e.g., if VS2 starts to sell an analogous mapping feature as an add-on option.

#### ASRs

*Security:* Some LiDAR data are stored on VS3's and VS1's LHDs. Thus, data security and integrity must be considered from the beginning - adding security to an already developed system is hard. On the other hand, the LiDAR data stored by the vehicle suppliers are not as sensitive as the final maps, thus security remains moderately important.

*Efficiency:* Efficiency is not an ASR from Boliden's perspective; this is the concern of the vehicle suppliers. Operations in this scenario is shared by the supplier and Boliden, creating a complex mutual dependency.

*Reliability:* The vehicle suppliers are not used to running a software service that requires monitoring throughout the operation. In addition, inaccuracies in the solution leading to incorrect maps might not be as obvious to a vehicle supplier compared to individuals working in the mine, i.e., they are not physically in the mine to detect discrepancies. Consequently, reliability should be considered early and constitutes an ASR.

### *Business perspective*

#### Inter-company dependency

Boliden relies on the vehicle suppliers to keep offering the mapping feature as an add-on option. Also, changing from VS1 will be less likely, as Boliden has spent effort on a customized transformation from their specific LiDAR data format.

#### Upfront investment

Boliden pays extra to obtain the mapping feature as an add-on option when purchasing LHDs from VS3 and VS1. Furthermore, development of a software component for transformation of LiDAR data from VS1 is outsourced to Acme Consulting.

#### Running costs

VS3's more advanced mapping feature offer is provided against a subscription fee. On the other hand, Boliden does not pay much to maintain the internal software parts – although resources from Acme Consulting will still be required.

### Risks

Risks in the add-on scenario is shared among Boliden and the vehicle suppliers, both in terms of development costs (upfront investment) and operations costs. The risk related to the operation can be regulated in contracts. Even though more data is handled by the suppliers, the different suppliers cannot access each other's data. Hence, the risks related to data privacy are limited.

### Innovation platform

The vehicle suppliers are more likely to identify innovative use cases with LiDAR then Boliden, but mapping is not part of their core businesses. Hence, the innovation setup is likely better than in Scenario In-house, but unlikely to generate disruptive solutions.

We conclude the description of Scenario Add-on Option with the SWOT analysis below.

Table 3. Scenario Add-on Option SWOT analysis

| | |
|---|---|
| **Strengths**<br>• R&D risks and effort by vehicle supplier<br>• Suppliers have capability in terms of large R&D organization | **Opportunities**<br>• Boliden can try pilot solution with one vehicle supplier, then add more incrementally.<br>• Benefit from investments the supplier is doing for other companies than Boliden |
| **Weaknesses**<br>• Data privacy – risk of data leaking to competitors from supplier IT systems<br>• Some lock-in to supplier and constituting solutions<br>• Fragmented solution – each supplier will have their own solution | **Threats**<br>• Vehicle suppliers might never offer the feature<br>• Risk of solution being cancelled<br>• Different suppliers choose different solutions<br>• Little or no control of roadmap / time-plans |

### 5.3 Scenario Additional Service - External Hauling Company Sells Mapping as an Optional Service

In Scenario Additional Service, Boliden has already outsourced the hauling service to the external company Acme Mining (analogous to the current agreement with "Bennys Gräv AB"). Acme Mining has also developed the innovative solution to update maps of underground maps using LiDARs mounted on their LHDs – and offers this to Boliden as an additional service. Boliden signs an SLA specifying the quality of the mapping service, but do not care which vehicle suppliers' LHDs Acme Mining operates in the mine.

Figure 7 summarizes Scenario Additional Service. On top of the figure, Acme Mining operated LHDs from VS1, VS2, and VS3 (cf. A). The solution used by Acme Mining contains three different software components for data transformation (cf. B), corresponding to the different LiDAR data formats provided by LHDs from the three vehicle suppliers. All LiDAR data are stored by Acme Mining (cf. C), and then they are processed (i.e., filtered and adapted to the 3DM import format, cf D.) prior to transfer to 3DM operated by Boliden. Finally, the LiDAR data is imported by 3DM to update its 3D CAD models (cf. E) – repeatedly synchronized with the maps in Pos (cf. F).

Looking at the business perspective, Boliden makes only minor upfront investments to realize the SoS benefits. Instead, Acme Mining drives the innovation by providing the mapping as an additional service – not only to Boliden, but to the mining market in general. An SLA (cf. G) between Acme Mining and Boliden regulates frequency and quality of the map updates, and Boliden pays considerable running costs for the service. While the costs are high, Acme Mining takes all risks related to quality of service, and Boliden does not engage in the technical SoS implementation, e.g., whether cloud solutions are used within Acme Mining. In Scenario Additional Service, Boliden does not develop any parts of the SoS themselves. Regarding operations, Boliden can focus on the same constituent systems as today, i.e., 3DM and POS.

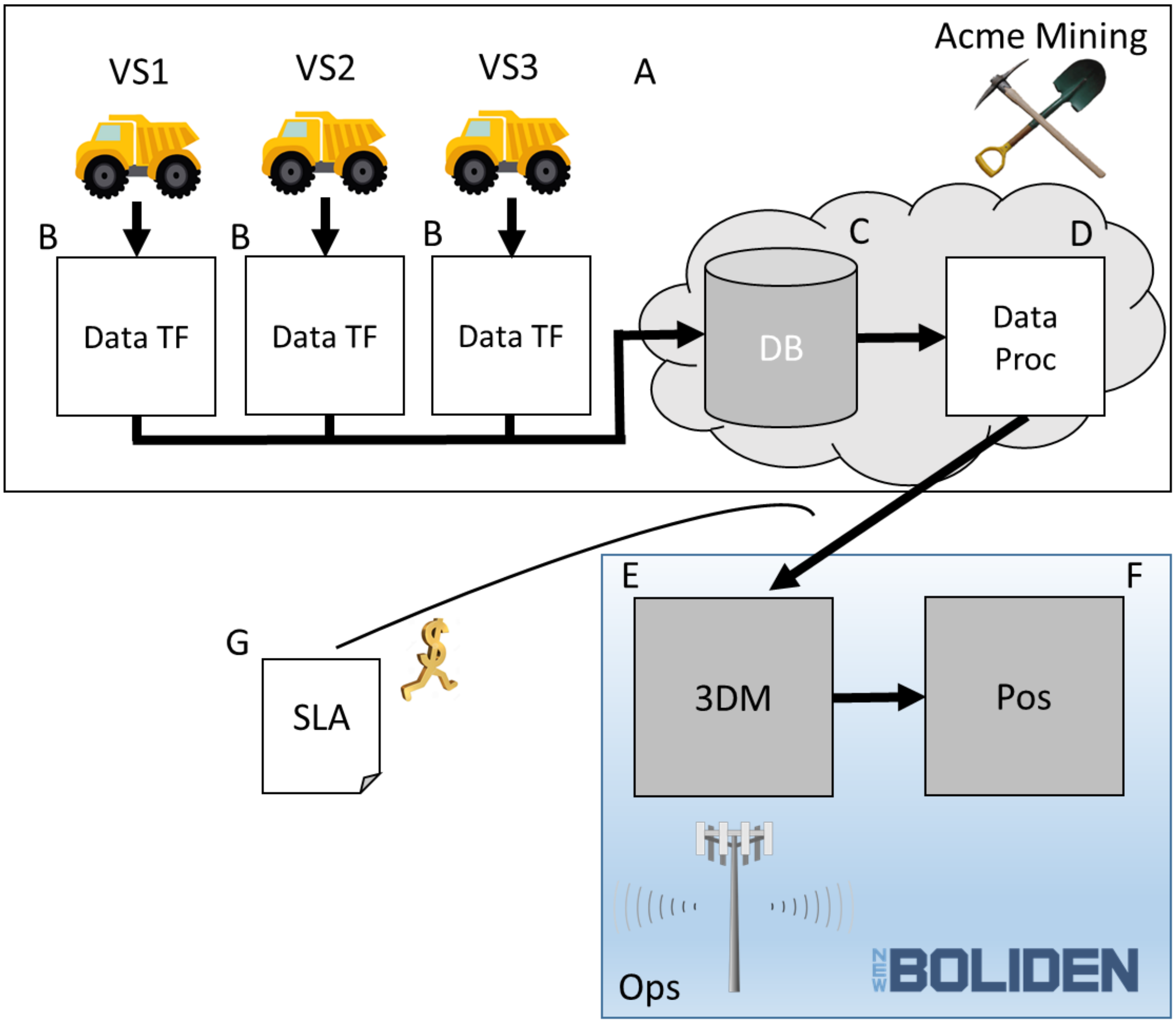

Figure 7: Overview of Scenario Additional Service. Acme Mining delivers the hauling service as regulated in an SLA, using LHDs from different vehicle suppliers. Boliden does no development.

Regarding the key variation points, we portray Scenario Additional Service as follows:

*Technical perspective*

### IT infrastructure

Scenario Additional Service outlines a solution where a hauling company not only drives the LHDs, they also operate a map rendering service based on the collected LiDAR data. The difference to Scenario Add-on Option is that the hauling company consolidates the data from different vehicle suppliers and acts as a single interface to Boliden. Presumably, the hauling company will use some kind of cloud solution.

### Evolution

Acme Mining offers mapping as a service to the general market, rather than a solution tailored for Boliden. Still, Boliden has the power to negotiate details of the LiDAR data transfer before signing a contract for the service. However, as the hauling company needs to align both suppliers and other customers, Boliden has less influence. The Boliden part in this SoS scenario is minimal. Most likely, all SoS evolution is driven by Acme Mining – possibly aligned with the vehicle suppliers' ongoing R&D. While it would be possible to bypass Acme Mining and obtain LiDAR data directly from some vehicle supplier, such a solution would require Boliden to invest in internal development, in line with Scenario In-house or Scenario Add-on option.

#### ASRs

*Security:* Acme Mining is responsible for the development and operations of the solution to render a map from LiDAR data. To achieve a scalable solution, they use the same cloud solution for all customers and do not have a separate data center for Boliden, i.e., the cluster for computation and storage is the same across their clients. As Acme Mining even renders the map, the final results more sensitive than just the raw LiDAR data. Thus, security in this scenario is of utmost importance.

*Efficiency*: As Acme Mining is responsible for the operation, efficiency is their responsibility and hence not an ASR from Boliden's perspective. However, it is important that efficiency is part of the SLA to ensure an operation suitable for Boliden's need.

*Reliability*. Reliability follows the same line of reasoning as efficiency, i.e., it is not an ASR but an important concern in the SLA.

### *Business perspective*

#### Inter-company dependency

The relationship with Acme Mining is fundamental in this scenario – the SoS exists only as long as they offer the mapping service. Switching to a different hauling company risks being a large undertaking, both technically as well as business wise.

#### Upfront investment

Boliden needs only limited initial investments. The required technology development is instead mainly funded by Acme Mining.

#### Running costs

Acme Mining charges Boliden well over what applies to the other scenarios, i.e., the running costs are comparably high. On the other hand, there are minimal maintenance costs for the SoS part operated by Boliden.

#### Risks

The investment risks are highest for Acme Mining. Boliden has a single partner with which they contractually regulate the quality of service, transferring risks to the hauling company . The biggest risk for Boliden is connected to data privacy, i.e., leaking of sensitive information regarding the ore body and the mining operations.

### Innovation platform

Acme Mining hopes to sell additional services to Boliden (and other customers), thus they work in close collaboration with several vehicle suppliers. Hence, they have a good opportunity to be innovative. However, they are not primarily in the mapping business, thus will not prioritize innovation in this area.

We conclude the description of Scenario Additional Service with the SWOT analysis below.

Table 4. Scenario Additional Service SWOT analysis

| | |
|---|---|
| **Strengths**<br>• R&D risk and cost with hauler<br>• A unified solution for Boliden | **Opportunities**<br>• Changes e.g. from vehicle suppliers will be handled by the hauler<br>• Fewer systems/partners to negotiate with |
| **Weaknesses**<br>• Data privacy<br>• Some lock-in to hauler – switching hauler solution can be difficult | **Threats**<br>• Acme Mining might never offer the service, or choose to discontinue it.<br>• Hauler might lack development resources → slow to handle updates and changes<br>• Boliden might be forced to adapt systems to fit the hauler<br>• The hauler might lack the competence (in-house) to do the development |

### 5.4 Scenario Ecosystem - An Open Software Ecosystem Enables a Third-party Mapping Solution

The final scenario we describe entails a solution that opens up the business and technical ecosystem. In Scenario Ecosystem, Boliden has an open technical platform through which services communicate and share data. While we present only the realization of the mapping service, the strength of an ecosystem is rather its facilitation of diverse services delivered by different providers. A software ecosystem implies openness and a common platform. This can for example mean sharing code, having open APIs or jointly working on a roadmap. There is no such thing as a completely open ecosystem, even though some open source projects come very close. In the Boliden case, some parts would be open but not others. For example, providing a platform with open APIs and SDK to selected partners. However, governance and control would likely not be opened up.

Figure 8 depicts an overview of Scenario Ecosystem. The essential part of the ecosystem is the open interface platform (cf. A), enabling communication within the SoS through the APIs of the constituent systems and software components – thus simplifying sharing of data and services, as well as integration of new parts. VS2 has an advanced LiDAR solution, including data processing and data storage (cf. B). The processed LiDAR data from VS2's LHDs are accessible, which they announce using the open interface platform (cf. C). Other service providers can then pull LiDAR data from VS2 for various purposes, e.g., map rendering. Moreover, VS3's LHDs provide access to raw LiDAR data, which other service providers can pull through the open interface (cf. D). In practice, Boliden pulls the data to a data processing component (cf. E), and then stores it in a database (cf. F), accessible for other service providers. In this Scenario, LHDs from VS1 do not offer any access to LiDAR data. 3DM recurrently updates its maps by importing LiDAR data through the open interface platform (cf. G), both from Boliden and VS2. Finally, Pos synchronizes with the latest spatial data (cf. H) as agreed upon with Boliden.

Figure 8 also shows key aspects from a business perspective. Boliden outsources development of the open interface platform, i.e., the central component facilitating all communication, to Acme Consultants (cf. Dev). In addition, Acme Consultants develops a data storage solution for Boliden to store LiDAR data from LHDs that do adhere to the 3DM import format. Boliden signs an agreement with VS3 to allow extraction of the LiDAR data, realized through upfront investments (cf. I), in line with Scenario In-house. All other players in the ecosystem are responsible for development of their own software, i.e, Acme Maps, VS2, and Pos, but Boliden pays subscription fees for access to services and their corresponding software maintenance.

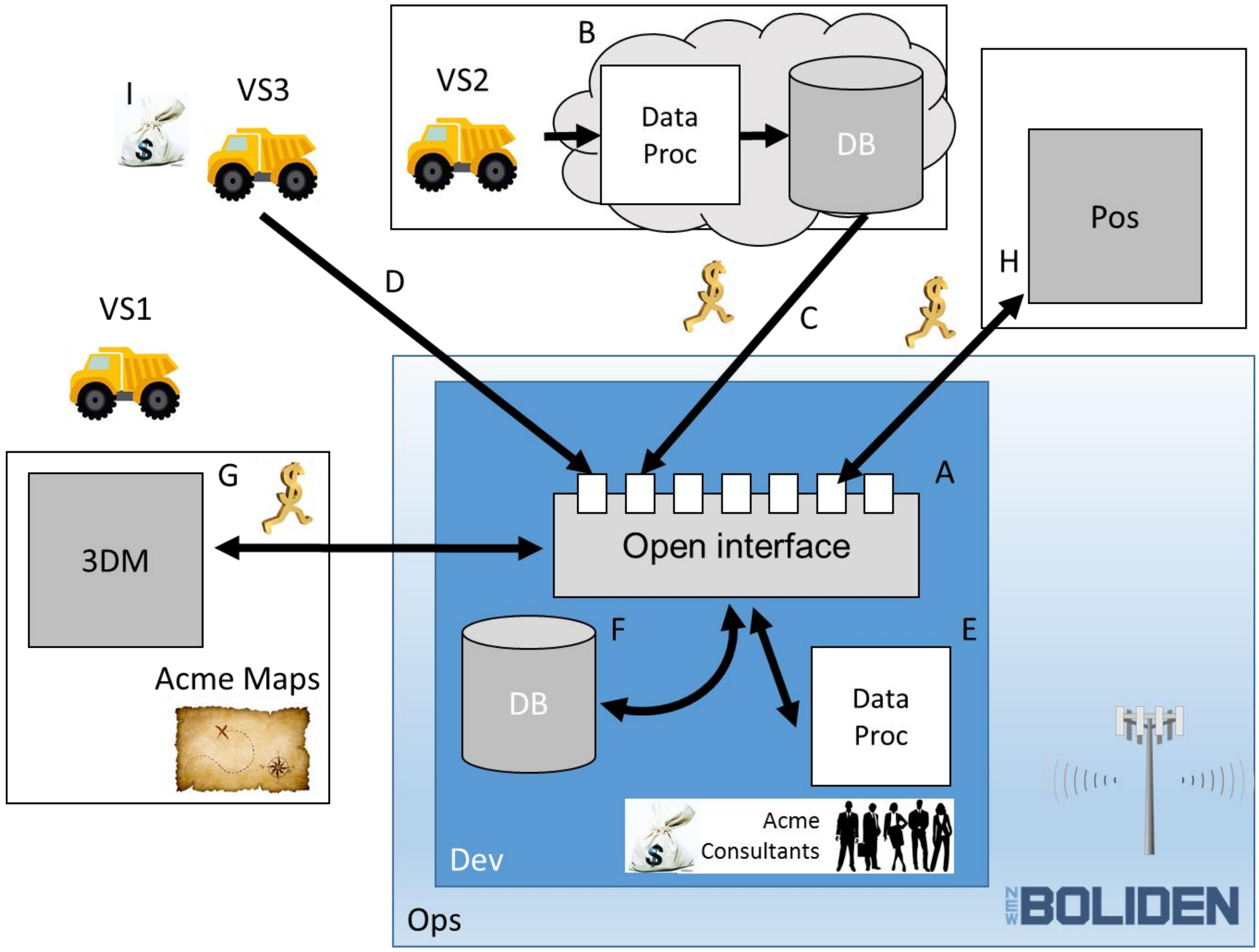

Figure 8: Overview of Scenario Ecosystem. Boliden facilitates a semi-open ecosystem that enables a third party mapping solution.

From an operations point of view, each partner is responsible for their respective parts of the ecosystem. Boliden outsources operations and maintenance of the open interface platform and the internal data storage to Acme Consultants. Compared to the other three scenarios, the emerging SoS is more collaboratively governed, with Boliden oversight though, and the different players are more independent. Although Boliden still has the overall ownership of the SoS, they need less bespoken development. Rather, they would rely on the initiatives and innovation of the different stakeholders and their own capability to provide off the shelf solutions for Boliden.

Regarding the key variation points, we characterize Scenario Ecosystem as follows:

*Technical perspective*

### IT infrastructure

In Scenario Ecosystem, the infrastructure is spread among the various partners. Boliden, with the support from Acme Consultants, operates and maintains the underlying open interface platform. The various partners in the ecosystem all operate their own solutions on their own IT infrastructure.

### Evolution

Compared to traditional development, evolution is much more dynamic in an ecosystem. Each partner continuously develops their solution, and the SoS composition changes over time in a flexible manner. Boliden can select the most suitable service from the available solutions in the ecosystem. However, tailoring to Boliden's specific needs is not as obvious – rather, Boliden must incentivize the ecosystem's different players to adapt.

### ASRs

In an ecosystem, different systems are interacting in a much more flexible and dynamic manner compared to traditional systems. Hence, *security* is vital to ensure that the right information is shared with the right system. Furthermore, having an open system where you beforehand do not know exactly which systems will interact with other systems, *efficiency* and *reliability* must be designed into the common open interface platform. If the common platform is not designed properly, then it does not matter how the partaking systems are implemented. However, if the platform is designed correctly, then it is up to each participating system/component to implement a sufficient level for the ASRs, of course in accordance with business agreements.

*Business perspective*

### Inter-company dependency

Integration of new equipment and services in Scenario Ecosystem is easy, almost seamless. An ecosystem opens up the possibility to easily change service providers. Hence, there is only minor low lock-in to specific suppliers and solutions.

### Upfront investment

Boliden makes an upfront investment in the open interface platform supporting the ecosystem, covering both the infrastructure and development outsourced to Acme Consultants. Also, development of the internal data storage and the data processing component constitute upfront investments.

### Running costs

Boliden must maintain the infrastructure around the open interface platform. The main running costs depend on actual usage of services from ecosystem partners, i.e., VS2, Acme Mining, and Pos. For example, if Boliden for some reason needs to cut costs, a service can be cancelled to save money – on the other hand, upfront investments might be lost.

### Risks

Investments and operations risks are shared among several partners. However, for the emerging properties, there is no single partner to agree on a quality of service. Consequently, if there is a need to regulate the service in contracts, it requires negotiations with several different companies. Data privacy is also a risk, as several partners have access to data access and no external partner feels responsible for the integrity of the complete solution – Boliden must take this role.

### Innovation platform

An argument for an open ecosystem is that open innovation is supported. In essence, as the playing field is open, anyone can come up with a new innovative idea and efficiently make it available to Boliden. It will not be Boliden who is driving the innovation, rather the partners.

We conclude the description of Scenario Ecosystem with the SWOT analysis below.

Table 5. Scenario Ecosystem SWOT analysis

| **Strengths** | **Opportunities** |
|---|---|
| • Shared R&D costs and risk among the ecosystem partners<br>• Utilizing innovation of the partner network.<br>• Maintenance and operation of parts of the IT systems will be handled by partners.<br>• Scalable solution – upscaling and downscaling easier than in the other scenarios | • Flexibility, easy to switch solution → better negotiation on price<br>• Faster and better innovation and improvements in the mine. |
| **Weaknesses** | **Threats** |
| • Upfront investment<br>• Sensitive data will be available to a wider audience, which need to be controlled by contracts.<br>• Quality assurance and operations of the emerging properties more challenging. | • Fail to attract interest for the ecosystem<br>• Technically more complex solution<br>• Partners might not be able to or willing to operate in a Boliden ecosystem.<br>• Solutions might not be exactly what Boliden wants, with less possibilities to influence. (Boliden is too small) |

## 6. Discussion

In this section, we discuss the results from our investigation. All four scenarios discussed in the previous section have their individual pros and cons. Table 6 shows a side-by-side comparison of the four scenarios from the perspective of the eight key variation points. Based on Table 6, we identify four aspects that are particularly important to consider when realizing a Kankberg SoS; We believe that cost, control, risk, and innovation constitute central elements in trade-offs that must be balanced in future work.

Table 6. Comparison of the four system-of-systems scenarios based on the key variation points.

| | **Scenarios** | | | |
|---|---|---|---|---|
| | **In-house** | **Add-on Option** | **Additional Service** | **Ecosystem** |
| **Technical perspective** | | | | |
| IT infrastructure | Centralized to Boliden | Shared with suppliers, consolidated by Boliden | Centralized around the hauling company | Decentralized |
| Evolution | Costly and slow | Supplier control | Hauler control | Dynamic |
| ASRs | Efficiency | Reliability | Security | All |
| **Business perspective** | | | | |
| Inter-company dependency | High | Medium | High | Low |
| Upfront investment | High | Low | Low | Medium |
| Running costs | High | Low | High | Medium |
| Risk | Focus on Boliden | Regulated in contract | Regulated in contract | Shared/unclear |
| Innovation platform | Weak | Medium | Medium | Strong |

We argue that four aspects are fundamental for discussions on how to proceed with a SoS solution in Kankberg: control, cost, risk, and innovation. Some of them are contradictory in nature, thus balancing trade-offs will be inevitable.

1. Control – Being able to control technical roadmaps, release plans, and quality levels is typically an advantage. If a particular SoS aspect is critical, and Boliden has very specific requirements both on content and plan, then control should be prioritized. There might also be non-technical aspects where Boliden might have specific requirements on, such as customer services or 24-hour support to handle any disturbances on the operations. That might also require more control from the Boliden side where a more loose cooperation in a SECO might not suffice. On the other hand, if Boliden does not have specific requirements and anyway not the means to invest, then giving up control is not necessarily a bad thing. In Table 6, there tends to be more control of the technical plans on the left-hand side. To some degree, control means "do it yourself", whereas if you want shared costs and/or innovation from outside Boliden, trust in others is essential.

2. Costs – Costs are tightly connected to time and quality level, as well as compromises in the functionality. Discussions on costs encompass both the total costs and upfront costs vs. running costs. Sharing costs with suppliers or even competitors means compromising. In a software ecosystem, the basic idea is essentially a specialization of actors and a sharing of basic costs [11]. In Table 6, the in-house scenario has the highest upfront cost as well as running costs. It is also the scenario in which Boliden has the highest control, hardly anything is shared with others. In terms of R&D cost, the two Scenarios Add-on Option and Additional Service are potentially the cheapest. On the other hand, the two scenarios provide less control of what and when the suppliers can deliver. A software ecosystem is short-term investment heavy, but can, if successful, be an economical option mid- and long-term.
3. Risk – The main business risks are related to the supplier network and their ability to deliver products and services. In Scenario In-House, Boliden carries all the risks as the overall integrator. Hence, both changes in technology and business climate would have to be handled by Boliden. Most importantly, there is a large financial risk as the upfront investment cannot be paid back. In the other scenarios, a possibility to cancel services remains. For Scenario Add-on Feature and Scenario Additional Service, on the other hand, there are risks related to the dependencies to other companies. There might be lock-in effects concerning either vehicle suppliers or the hauling company. Lock-ins increase the risk of price increases, as switching partners is more difficult when solutions are unique to the partners. Even though some risks can be handled contractually in an SLA, writing good SLAs are difficult and not necessarily in the business interest of the various stakeholders. In Table 6, the two scenarios in the middle are the ones with the lowest risks. Scenario Ecosystem is unique, as risks are on the one hand shared in the ecosystem, but, on the other hand, there is little or unclear control. Hence, risk management becomes a matter of trust among the actors in the software ecosystem.
4. Innovation – Boliden is in the mining business and not in the digital innovation business. However, the digitalization of society will lead to innovation everywhere. Even if Boliden is not a technical innovation company, innovation will happen in the company's use cases and business models. Hence, Boliden will also need a foundation for a successful innovation platform. Various types of data are becoming more and more important for innovation in general and new business models in particular. Hence, special attention should be put on understanding both the value of data as well as risks of sharing data. In Table 6, the scenarios on the right-hand side, enabling an environment of open innovation [17], appear to provide a better innovation platform than the scenarios on the left-hand side. Especially, the in-house scenario will suffer as the ideas must primarily come from Boliden and not so much from outside influence and new perspectives.

Thorough analysis and pre-studies are required to identify how to best balance the trade-offs regarding a Kankberg SoS; any decision must be aligned with Boliden's strategic goals. It is evident that a customized fully controlled in-house solution will decrease the potential for open innovation, and also be costly – even more so if risks need to be minimized. On the other side of the spectrum, Scenario Ecosystem strongly supports innovation, but the risk that there will be no mature actors in the software ecosystem is high, and also out of Boliden's control. Looking at Scenario Add-on Option and Scenario Additional Feature, these scenarios require less upfront investments, but Boliden is not in control of when such a SoS could be available on the market. Finally, we want to highlight that the four scenarios are not mutually exclusive; a Kankberg SoS could be realized through a combination of in-house solutions, add-on options purchased from vehicle suppliers, and services bought from hauling companies – such a variety of solutions could even co-exist in a software ecosystem.

## 7. Summary and Concluding Remarks

Ever-increasing production targets under strict safety requirements pushes mining operations toward increased automation. In the not so very distant future, fewer people will work underground – instead, mining equipment such as LHD trucks will be remote controlled or even autonomous. As there is no GPS signal in an underground mine, the LHD trucks will navigate using LiDAR. The deployment of LHDs equipped with LiDAR sensors will result in considerable amounts of detailed spatial data, i.e., point clouds from 3D scanning, collected as the LHD trucks drive through drifts and crosscuts.

In the PIMM project, a novel approach to establish reliable wireless communication in the Kankberg mine is explored. By constructing an underground 5G network, high bandwidth wireless communication would be possible in an environment previously restricted to wires or locally managed Wi-Fi networks. Given a reliable wireless communication network in Kankberg, several new opportunities arise; Also in the harsh setting of an underground mine, software-intensive solutions relying on high data transfer rates could ripe the benefits of increased digitalization, i.e., embracing the predicted data abundance.

In this report, we illustrated one possible future approach to utilizing reliable wireless communication: a system-of-systems in the Kankberg mine. We analyzed how LHD haulers equipped with LiDAR could be organized into a system-of-systems together with existing map rendering software and indoor positioning systems. Such a solution could use LiDAR data collected from LHD trucks to continuously update a detailed 3D map of the Kankberg mine. The primary benefit would be related to safety of the mining operations, i.e., detecting potential cave-ins, but it would also contribute to productivity. A highly accurate map updated in near real-time would be valuable input to support maintenance activities, and in general to overview ongoing mining operations.

In our analysis, we presented four different scenarios and business models that could enable the Kankberg system-of-systems: 1) an in-house solution, 2) buying mapping capability as an add-on feature from the vehicle suppliers, 3) paying for mapping as an additional service from the hauling service provider, and 4) a software ecosystem. For each scenario, we discussed eight key variation points covering both technical and business perspectives, and we presented corresponding SWOT analyses. We discussed how the key variation guided us to inevitable trade-off discussions concerning control, costs, risks, and innovation. An in-house solution would provide full control of features and risks, but be costly and provide a weaker innovation platform.

Based on our analysis, we propose three ways forward:

- LiDAR mapping feasibility study – even though the LiDAR mapping functionality is mainly a means to stimulate SoS discussions in this report, it would be interesting to set up a pilot study to explore its feasibility. It should include not only the technical challenges but also the potential partners to work through a full-fledged business scenario.
- Software ecosystem pre-study – for a software ecosystem to thrive, it requires both a robust technical platform as well as a group of actors, i.e., a community. Would it be realistic for the future of the mining industry to use a software ecosystem paradigm in the IT infrastructure and innovation platform? A pre-study involving potential actors and their outlooks could analyze this in more detail, with a focus on open innovation.

- System-of-systems deep-dive – in a SoS, operation and control are different compared to traditional systems. It is not obvious how to set up the IT infrastructure and enterprise architecture to ensure efficiency, quality, and maintainability. An in-depth study on the current IT infrastructure, combined with elicitation of infrastructure requirements needed to enable a SoS, could provide necessary answers on what needs to be changed in Kankberg.

Analogous to other traditional industries, the future of mining will be increasingly dependent on software. All future projections point toward digitalization, shared data, and increased automation. This new environment will fundamentally change business models, and software will be the main driver in R&D investments. Given reliable wireless communication, software will be the glue that enables systems-of-systems in mines such as Kankberg – the question is, who takes the lead?

## Acknowledgements

This work was funded by VINNOVA, the Swedish Agency for Innovation Systems within the PIMM project, Pilot for Industrial Mobile communication in Mining.